\documentclass[prl,twocolumn,showpacs,floatfix]{revtex4}

\usepackage{graphicx}
\graphicspath{{./Eps/}}
\usepackage{amsmath}
\usepackage{amssymb}

\begin{document}

\title{Note on the Coulomb blockade of a weak tunnel junction
with Nyquist noise: Conductance formula for a broad temperature range}

\author{%
  Ant\'{o}nia Mo\v{s}kov\'a\textsuperscript{\textsf{\bfseries 1}},
  Martin Mo\v{s}ko\textsuperscript{\textsf{\bfseries 1},\textsf{\bfseries 2},}}

\email{martin.mosko@savba.sk}

\affiliation{
  \textsuperscript{1}\,Institute of Electrical Engineering, Slovak
  Academy of Sciences, 841 04 Bratislava, Slovak Republic\\
  \textsuperscript{2}\ Faculty of Mathematics, Physics and Informatics, 
  Comenius University, Mlynsk\'a dolina, 84248 Bratislava, Slovak Republic}


\begin{abstract}
We revisit the Coulomb blockade of the tunnel junction with conductance much smaller than $e^2/\hbar$. 
We study the junction with capacitance $C$, 
embedded in an Ohmic electromagnetic environment modelled by a series resistance $R$ which produces the Nyquist noise. 
In the semiclassical limit the Nyquist noise charges the junction by a random charge with a Gaussian distribution. 
Assuming the Gaussian distribution, we derive analytically the temperature-dependent junction conductance 
$G(T)$ valid for temperatures $k_BT \gtrsim (R_K/2\pi R)E_c$ and resistances 
$R \gtrsim R_K$, where $R_K = h/e^2$ and $E_c=e^2/2C \ \text{is}$
the single-electron charging energy. 
Our analytical result shows the leading
dependence $G(T) \propto e^{-E_c/4k_BT}$, so far believed to exist only 
if $(R_K/\pi R)E_c \ll k_BT \ll E_c$ and $R \gg R_K$. 
The validity of our result for $k_BT \gtrsim (R_K/2\pi R)E_c$ and $R \gtrsim R_K$ 
is confirmed by a good agreement with the numerical studies which do not assume 
the semiclassical limit, and by a reasonable agreement with experimental data for $R$ as low as $R_K$.
Our result also reproduces various asymptotic formulae derived in the past.
The factor of $1/4$ in the activation energy $E_c/4$ is due to the semiclassical Nyquist noise.
\end{abstract}

\maketitle
\section{I. Introduction}

Electron tunneling in a small tunnel junction is affected by
the electromagnetic environment which gives rise to the Coulomb blockade.
Since the resulting current-voltage characteristics
depends on the type of the
environment, one speaks about a coupled
junction-environment  system \cite{Averin,Delsing,NazarovJPT,SchoenZaikin,IngoldNazarov}.
Starting from these ideas,
the Coulomb-blockaded current-voltage characteristics of the coupled
junction-environment system was derived for a so-called weak tunnel
junction \cite{DevoretPRL,GirvinPRL} 
which has the tunnel conductance much smaller than $e^2/h$. 
Later, the theory was extended for a junction with arbitrary strong tunneling \cite{JoyezPRL}.
The theory \cite{DevoretPRL,GirvinPRL,JoyezPRL} gives a formal result
for the current voltage characteristics, however, analytical results are known
\cite{Averin,AverinBook,DevoretPRL,GirvinPRL,GerdSchoen,GrabertNATOARW,Ingold94,IngoldHabilitation,Heikilla,Panyukov,Kauppinen,Wang,JoyezPRB} 
only for a special limits and a full comparison 
of the theory and experiment \cite{JoyezPRL} requires a numerical calculation \cite{JoyezPRB,Odintsov}.
In particular, a temperature-dependent junction conductance, $G(T)$,
was derived analytically
\cite{Averin,AverinBook,Panyukov,Kauppinen,Wang,JoyezPRB} either for $k_BT \gg E_c$ or for $k_BT \ll E_c$,  
where $E_c=e^2/2C$ is the single-electron charging energy of the junction with capacitance $C$. 
As far as we know,  for the temperature range from $k_BT \gg E_c$ down to $k_BT \ll E_c$ 
only the numerical studies of $G(T)$ are known \cite{GirvinPRL,JoyezPRL,JoyezPRB,Flensberg}.

Here we revisit the Coulomb blockade of the weak tunnel junction with the aim to derive analytically $G(T)$ 
for a broad temperature range. 
We consider the junction with capacitance $C$, embedded in
an electromagnetic environment modeled by a series resistance $R$ which produces the Nyquist noise. 
In the semiclassical limit the Nyquist noise charges the junction by a random charge with a Gaussian 
distribution \cite{IngoldNazarov,GerdSchoen,Heikilla}. 
Assuming the Gaussian distribution, we derive an analytical $G(T)$ expression
valid for $k_BT \gtrsim (R_K/2\pi R)E_c$, where $R_K = h/e^2$ and $R \gtrsim R_K$.

Our $G(T)$ expression shows the leading dependence $G(T) \propto e^{-E_c/4k_BT}$,  so far known 
\cite{Averin,Odintsov} to exist only in conditions $(R_K/\pi R)E_c \ll k_BT \ll E_c$ and $R \gg R_K$. 
The validity of our result for $k_BT \gtrsim (R_K/2\pi R)E_c$ and $R \gtrsim R_K$ 
is confirmed 
by a good agreement with the numerical studies \cite{GirvinPRL,JoyezPRL,JoyezPRB,Flensberg} 
which do no rely on the semiclassical limit, and by a reasonable agreement with experimental data  
\cite{JoyezPRL} for $R \simeq R_K$.
Our result also reproduces various asymptotic formulae derived in the past \cite{Averin,Wang,JoyezPRB}.
Finally, we point out that the factor of $1/4$ in the activation energy $E_c/4$ is due to the 
Nyquist noise and we add a simple treatment of the noise in the semiclassical limit.

Sect. II starts with a review of the weak tunnel junction theory \cite{DevoretPRL,GirvinPRL} 
which serves as our starting point.
Then we derive  $G(T)$. In Sect.III our results are discussed and compared with previous works.
A summary is given in Sect. IV and we finish with an appendix which discusses the Nyquist 
noise within a semiclassical transport model.
This paper is a refined version of our recent arXiv attempts \cite{arXiv}.

\section{II. Theoretical considerations}

We first briefly review the weak tunnel junction theory \cite{DevoretPRL,GirvinPRL}.
Figure 1a shows two electron gases separated by energy barrier $V_0$.
Assume that the electrons pass through the barrier only by tunneling.
The junction conductance without the Coulomb blockade, $G_t$,
obtained from a golden-rule approach, reads
$G_t = (4 \pi e^2/\hbar) N_l N_r \mid \mathcal{T} \mid^2$,
where $\mathcal{T}=<r \mid H_t \mid l>$ is the matrix element
(approximated by a constant) for tunneling from an initial state $\mid l>$
on  the left hand side of the barrier to a final state $\mid r>$ on the right hand side,
$N_{l/r}$ is the density of states at the Fermi energy on the left/right hand side,
and $H_t$ is the weakly-perturbing tunneling Hamiltonian \cite{FerryGoodnick,Pfannkuche}.

In the electric circuit of figure 1b the junction 
with capacitance $C$ and tunneling conductance $G_t$ is coupled to a constant voltage source $V_x$
via the impedance $Z(\omega)$ which represents the impedance of the external circuit.  
This impedance, defined as $Z(\omega) = U(\omega)/I(\omega)$,  
gives the ratio between an alternating voltage of frequency $\omega$ and the current 
which is flowing through it if the junction is replaced by a short 
\cite{IngoldNazarov,DevoretPRL,GerdSchoen,Heikilla}. 
Due to the coupling to the external circuit the voltage drop at the junction fluctuates 
but its average value $V$ fulfills the Kirchhoff law \cite{GerdSchoen}
\begin{equation}
V + I(V)Z(0) = V_x
\ ,
\label{Kirchhoflaw1}
\end{equation}                 
where $I(V)$ is the current-voltage characteristic.
The junction can be reduced to a capacitor if $1/G_t \gg Z(\omega)$ \cite{GerdSchoen,Heikilla}. 
In such case the total impedance at the site of the junction is  \cite{GerdSchoen,Heikilla}
 $Z_t(\omega) = 1/[i\omega C + 1/Z(\omega)]$.

\begin{figure}[!t]
\centerline{\includegraphics[clip,width=0.9\columnwidth]{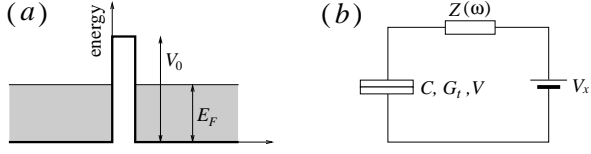}}
\caption{
(a) Conduction band edge in a tunnel
junction due to two weakly linked metals.
(b) Tunnel junction in the circuit.
}
\label{FIGURE1}
\end{figure}

Further, $I(V)= e[\Gamma^{+}(V) - \Gamma^{-}(V)]$,
where \cite{GerdSchoen,IngoldHabilitation}
\begin{equation}
 \Gamma^{+}(V) = \frac{G_t}{e^2} \int_{-\infty}^{\infty} dE
\frac{E}{1-\exp{(-E/k_BT)}} P(eV-E)\
\label{Tunnelingrate}
\end{equation}
is the rate of tunneling from the left reservoir to the right one,
$\Gamma^{-}(V) = \Gamma^{+}(-V)$ is the tunneling rate in the opposite direction,
and function $P(E)$ is
the probability that a tunneling electron creates an environmental excitation
with energy $E$.
The above equations give
\cite{DevoretPRL,GerdSchoen,JoyezPRB}
\begin{equation} \begin{split}
 I(V)  &
 = \frac{G_t}{e} \int_{-\infty}^{\infty} dE \frac{E}{1-\exp{(-E/k_BT)}}    \\
     &
\times   [P(eV-E)-P(-eV-E)]
\ .
\end{split}
\label{IVexpressionviaPE}
\end{equation}
Function $P(E)$ is related \cite{DevoretPRL,GerdSchoen,JoyezPRB}
to the correlation function $J(t) = <[\phi(t)-\phi(0)]\phi(0)>$
of the phase $\phi(t)=\int_{-\infty}^{t}dt [eU(t) - eV]$ across the impedance
$Z_t$. Specifically,
\begin{equation}
P(E) = (1/2 \pi \hbar) \int_{-\infty}^{\infty} dt \exp{[J(t)+i E t/\hbar]}
\ .
\label{PEorthodoxtheory}
\end{equation}
Finally, from the fluctuation dissipation theorem one finds
\begin{equation}
J(t)
= 2 \int_{-\infty}^{\infty} \frac{d\omega}{\omega}
\frac{\mathrm{Re}{\{Z_t(\omega)\}}}{R_K}
\frac{\exp{(-i\omega t)}-1}{1 - \exp(-\frac{\hbar \omega}{k_BT})} \  ,
\label{Jtorthodoxtheory}
\end{equation}
where $R_K \equiv h/e^2$ is the resistance quantum
\cite{DevoretPRL,GerdSchoen,IngoldHabilitation,JoyezPRB}.

The above theory is assumed to hold if $H_t$ is a weak perturbation and $1/G_t \gg Z(\omega)$.
Moreover, it has to be fulfilled that
$1/G_t \gg R_K$, which ensures \cite{Korotkov} that
a single-electron tunnel event takes a much shorter time than the time
between two events.
The electrons are thus mostly localized on the electrodes, which is an implicit
assumption of the model in figure 1b. 
We will see later that the condition $1/G_t \gg R_K$ 
can be soften to \cite{Heikilla} $1/G_t \gg R_K/2 \pi$ 
and condition $1/G_t \gg Z(\omega)$ is likely too stringent as well.

We consider the Ohmic
environment \cite{GerdSchoen,Heikilla} $Z(\omega)=R$, for which
$\mathrm{Re}{\{Z_t(\omega)\}} = R/(1+\omega^2 R^2 C^2)$.
For $R \gg R_K$ 
one has $\mathrm{Re}{\{Z_t(\omega)\}} = (\pi/C) \delta(\omega)$ and
one obtains \cite{GerdSchoen,Heikilla} the result $J(t) = - \pi/ (CR_K) [it+k_BTt^2/\hbar]$ 
valid \cite{GerdSchoen} for $\hbar/RC \ll k_BT$.
Setting the last $J(t)$ expression into the equation \eqref{PEorthodoxtheory}
one finds \cite{GerdSchoen,Heikilla}
\begin{equation}
P(E) =  \frac {1}{\sqrt{4\pi k_bTE_c}}
 \exp{\left[-\frac{(E-E_c)^2}{4k_bTE_c}\right]}, \   k_BT \gg \frac{\hbar}{RC},
\label{PEorthodoxtheoryGaussian}
\end{equation}
where $E_c \equiv e^2/2C$.
This limit is called semiclassical and the Gaussian spread of the distribution 
\eqref{PEorthodoxtheoryGaussian}
is due to the Nyquist noise produced by resistance $R$.
To see all this in a simple way, in the appendix we derive the Nyquist noise and
probability \eqref{PEorthodoxtheoryGaussian} from a semiclassical transport model.

Now we derive the conductance $G$ 
which relies on the approximation \eqref{PEorthodoxtheoryGaussian}. 
Thus, our derivation should be valid only for $k_BT \gg \hbar /RC$ and $R \gg R_K$. 
However, we will see that the obtained conductance result works quite well 
down to $k_BT \simeq \hbar/2RC$ and $R \gtrsim R_K$.

We rewrite equation \eqref{IVexpressionviaPE} for $V \rightarrow 0$ as
\begin{equation}
I(V)
= G_t \int_{-\infty}^{\infty} dE \frac{2E}{1-\exp{(-E/k_BT)}} \frac{dP(-E)}{d(-E)}V
\ .
\label{IVexpressionviaPEder}
\end{equation}
Replacing $-E$ by $E$ and inserting for $P(E)$ the equation
\eqref{PEorthodoxtheoryGaussian} we obtain the Ohm law
$I = GV$, where
\begin{equation} \begin{split}
G    &
=  \frac {G_t}{\sqrt{4\pi k_bTE_c}} \int_{-\infty}^{\infty} dE
\frac{2E}{\exp{(E/k_BT)}-1}  \\
   &
\times    \
\frac{d}{dE}  \left(\exp{\left[-\frac{(E-E_c)^2}{4k_bTE_c}\right]}\right) \ ,
\ \ \ \ \ k_BT \gg \frac{\hbar}{RC} \ .
\end{split}
\label{ConductanceexpressionviaPEderGaus}
\end{equation}
The integral in equation \eqref{ConductanceexpressionviaPEderGaus}
cannot be calculated analytically.
However, we can use a simple trick which allows us to asses $G(T)$ for all considered $T$
without such calculation.
Using $x = E/2k_BT$ and $\alpha \equiv E_C/k_BT$ we get
\begin{equation} \begin{split}
\frac{G}{G_t}   &
=  \frac{1}{\sqrt{\pi \alpha}} \int_{-\infty}^{\infty} dx
\frac{2x}{\exp{(2x)}-1}  \\
   &
\times    \
\frac{d}{dx}  \left(\exp{\left[-\left(\frac{x}
{\sqrt{\alpha}}-\frac{\sqrt{\alpha}}{2}\right)^2 \right]}\right)
\ ,
\end{split}
\label{ConductanceexpressionviaPEderGaussubst}
\end{equation}
where $1/\alpha \gg R_K/\pi R$.
After a simple calculation we get
\begin{equation}
\frac{G}{G_t} = \exp{\left(-\frac{\alpha}{4}\right)} I({\alpha}) \ ,
\ \ \ \ \ \ \ \ \frac{1}{\alpha} \gg \frac{R_K}{\pi R}
\ ,
\label{conductivity1}
\end{equation}
where
\begin{equation}
I(\alpha) = \frac{1}{\sqrt{\pi\alpha}}
\int_{-\infty}^{\infty}dx
\exp{\left(-\frac{x^2}{\alpha}\right)} \frac {x}{\sinh(x)}
\ ,
\label{correction1}
\end{equation}
or alternatively, by means of variable $y = x/\sqrt{\alpha}$,
\begin{equation}
I(\alpha) = \frac{1}{\sqrt{\pi}}
\int_{-\infty}^{\infty}dy
\exp{\left(-y^2 \right)} \frac {\sqrt{\alpha }y}{\sinh(\sqrt{\alpha} y)}
\ .
\label{correction2}
\end{equation}
Equation \eqref{conductivity1} is a useful alternative form of
equation \eqref{ConductanceexpressionviaPEderGaus}.

From equation \eqref{conductivity1} we obtain the leading dependence
\begin{equation}
\frac{G}{G_t} \simeq \exp{\left(-\frac{\alpha}{4}\right)} =
\exp{\left(-\frac{E_c}{4k_BT}\right)} \ ,
\ \ \ \ \frac{1}{\alpha} \gg \frac{R_K}{\pi R}
\ . \ \
\label{conductivityLeadingdep}
\end{equation}
Indeed, for $\alpha \rightarrow 0$ we find from equation \eqref{correction1}
result $I(0)=1$ while for $\alpha \gg 1$
we find
\begin{equation}
I(\alpha) \simeq \frac{1}{\sqrt{\pi\alpha}}
\int_{-\infty}^{\infty} dx \
\frac{x}{\sinh{x}}=\sqrt{\frac{\pi^3}{4\alpha}}
\ ,
\label{correction inftyB}
\end{equation}
because $\exp{(-x^2/\alpha)} \rightarrow 1$
for $\mid x \mid \lesssim \sqrt{\alpha}$
and $x/\sinh{(x)}$ is peaked at $x = 0$.
We see that $I(\alpha)$ decays with increasing $\alpha$
from $I(0)=1$  towards the weak dependence
 $I(\alpha) = \sqrt{\pi^3/4\alpha}$  at large $\alpha$.
Thus, the leading dependence in equation \eqref{conductivity1}
is indeed $\exp{(-\alpha/4)}$ for any $\alpha$.

To get beyond the leading dependence \eqref{conductivityLeadingdep}
we examine again the factor $I(\alpha)$. For $\alpha \ll 1$ we can use expansion
\begin{equation}
\frac {\sqrt{\alpha }y}{\sinh(\sqrt{\alpha} y)}
 = 1-\frac  {\alpha}{6} y^2+\frac {7\alpha^2} {360}  y^4+O[(\sqrt{\alpha} y)^6]
\ .
\label{expansion small}
\end{equation}
Inserting expansion \eqref{expansion small} into the equation \eqref{correction2}
we obtain
\begin{equation}
I(\alpha)
 = 1-\frac {\alpha} {12} +\frac {7\alpha^2} {480}  +O\left[\alpha^3\right]
\ , \ \ \ \ \ \ \alpha \ll 1 \ .
\label{correction2 small}
\end{equation}
For $\alpha \gg 1$ we can use expansion
\begin{equation}
\exp{\left( -\frac{x^2}{\alpha}\right)}
 = 1-\frac {x^2} {\alpha} + \frac {x^4} {2 \alpha^2}
+O\left[(x/\sqrt{\alpha})^6\right]
\ .
\label{expansion big11}
\end{equation}
Setting expansion \eqref{expansion big11} into the equation \eqref{correction1}
we obtain
\begin{equation}
I(\alpha) = \sqrt{\frac{\pi^{3}}{4\alpha}}
\left[
 1-\frac{1}{\alpha}\frac{\pi^2}{2}+\frac{1}{\alpha^2} \frac{\pi^4}{2}
 + O\left(\frac{1}{\alpha^3}\right)
 \right]
\ ,
\label{correction big22}
\end{equation}
where $\alpha \gg 1$. For large enough $\alpha$ the last equation reduces
to $I(\alpha) \simeq \sqrt{\pi^3/4\alpha}$,
in accord with limit  \eqref{correction inftyB}.
Finally, we propose a simple but very precise interpolation between
$I(0)=1$ and $I(\alpha \gg 1) = \sqrt{\pi^3/4\alpha}$, namely
\begin{equation}
I(\alpha)
\simeq \left(1+4\alpha/\pi^3\right)^{-1/2}
\ .
\label{interpolation}
\end{equation}

\begin{figure}[!t]
\begin{minipage}[t]{0.46\textwidth}
\begin{center}
\includegraphics[clip,width=1.\textwidth]{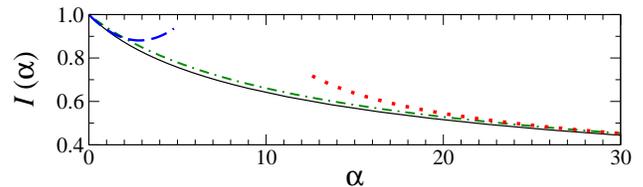}
\end{center}
\caption{
Factor $I(\alpha)$ versus $\alpha$.
The full line is obtained by calculating numerically the integral \eqref{correction1}.
Interpolation \eqref{interpolation} is shown in a dotted-dashed line.
Limits \eqref{correction2 small} and $\eqref{correction big22}$
are shown in a dashed line and in a dotted line, respectively.
}
\label{FIGURE2}
\end{minipage}
\end{figure}

One can see in figure 2 that the interpolation \eqref{interpolation}
fits the exact $I(\alpha)$ almost perfectly while expansions \eqref{correction2 small}
and $\eqref{correction big22}$ work only for a too small or too large $\alpha$.
Combining equations  \eqref{interpolation} and \eqref{conductivity1}
we obtain the result
\begin{equation}
\frac{G}{G_t} \simeq \left(1+\frac{4\alpha}{\pi^3}\right)^{-1/2}
\exp{\left(-\frac{\alpha}{4}\right)} \ ,  \ \ \ \ \ \
\frac{1}{\alpha} \gg \frac{R_K}{\pi R}
\ ,
\label{conductivity2}
\end{equation}
which goes beyond the leading approximation \eqref{conductivityLeadingdep}.
Since the interpolation \eqref{interpolation} works very well, result \eqref{conductivity2}
agrees with the exact result \eqref{conductivity1} almost perfectly.

Results \eqref{conductivity1}, \eqref{conductivityLeadingdep}, and \eqref{conductivity2}
did not yet appear in the literature. 
In the next section we will see that the restriction $1/\alpha \gg R_K/\pi R$
can be soften to $1/\alpha \gtrsim R_K/2\pi R$.

Noteworthy, $G/G_t \simeq \exp{(-E_c/4k_BT)}$ is the Arhenius dependence with
activation energy $E_c/4$ rather than $E_c$. We wish to point out that the factor 
of $1/4$ in the activation energy $E_c/4$ is due to the Nyquist noise.
First of all, as shown in the appendix, the Nyquist noise gives rise to the finite 
width of the distribution \eqref{PEorthodoxtheoryGaussian}.
If we ignore the noise and replace 
the distribution \eqref{PEorthodoxtheoryGaussian} 
by $P(E) = \delta(E-E_c)$, 
the equation \eqref{ConductanceexpressionviaPEderGaus} reduces to
\begin{equation} \nonumber
\frac{G}{G_t}
=  \int_{-\infty}^{\infty} dE
\frac{2E}{\exp{(E/k_BT)}-1}  \ \frac{d}{dE}  \delta(E-E_C) \\
   .
\label{ConductanceexpressionviaPEdelta}
\end{equation}
From the last equation one readily obtains (for $k_BT \ll E_c$) the leading dependence
$G \propto \exp{(-E_c/k_BT)}$. Unlike the dependence $G \propto \exp{(-E_c/4k_BT)}$, 
this leading dependence does not contain the factor of $1/4$.

The leading dependence $G \propto \exp{(-E_c/4k_BT)}$
has been derived already in Refs. \cite{Averin} and \cite{Odintsov}, 
however, only for
$(R_K/\pi R)E_c \ll k_BT \ll E_c$. We have just shown that it is
valid for $k_BT \gg (R_K/\pi R)E_c$ and
we will extend this range of validity even more in the next section.

\section{III. Discussion of results, comparison with previous works, asymptotic behavior}

We have derived the formulae \eqref{conductivity1}, \eqref{conductivityLeadingdep},
and \eqref{conductivity2} 
by considering the conditions $G_t^{-1} \gg R \gg R_K$ and $k_BT \gg (R_K/\pi R)E_c$.  
The last condition implies that the formulae are applicable in a broad temperature range 
(from $k_BT \gg E_c$ down to $k_BT \ll E_c$) only if $R$ exceeds $R_K$ 
almost two orders of magnitude. 
Since a practically feasible values of $R$ are $R \sim R_K$ \cite{JoyezPRL,Cleland}, 
it may seem that the formulae \eqref{conductivity1}, \eqref{conductivityLeadingdep},
and \eqref{conductivity2} are not applicable in practice. 
Surprisingly, we find below that they in fact work for temperatures 
$k_BT \gtrsim (R_K/2\pi R)E_c$ and resistances $R \gtrsim R_K$. 
This extends their applicability quite remarkably.

In figure 3 the formulae \eqref{conductivity1}, \eqref{conductivityLeadingdep},
and \eqref{conductivity2} are compared with
the numerical data \cite{JoyezPRB} which hold for any $R$. 
More precisely, the calculation \cite{JoyezPRB} is still restricted 
to a tunnel junction 
with $G_t^{-1} \gg R_K$ and $G_t^{-1} \gg R$, 
however, it is not restricted by assumptions $\pi R/R_K \gg E_C/k_BT$ 
and $R \gg R_K$
as it does not use the approximation \eqref{PEorthodoxtheoryGaussian}.

As expected, results \eqref{conductivity2} and \eqref{conductivity1} almost coincide.
Further, one can see that they agree with the numerical data \cite{JoyezPRB} 
for $R/R_K = 10$ in the whole temperature range. 
This is a surprising finding because for $R=10R_K$
the assumption $(R_K/\pi R)E_c \ll k_BT$ implies that the results \eqref{conductivity2} 
and \eqref{conductivity1} should deviate from the numerical data remarkably 
if $k_BT \lesssim E_c/ \pi$. 
However, there is no deviation at all.

\begin{figure}[!t]
\begin{minipage}[t]{0.46\textwidth}
\begin{center}
\includegraphics[clip,width=1.\textwidth]{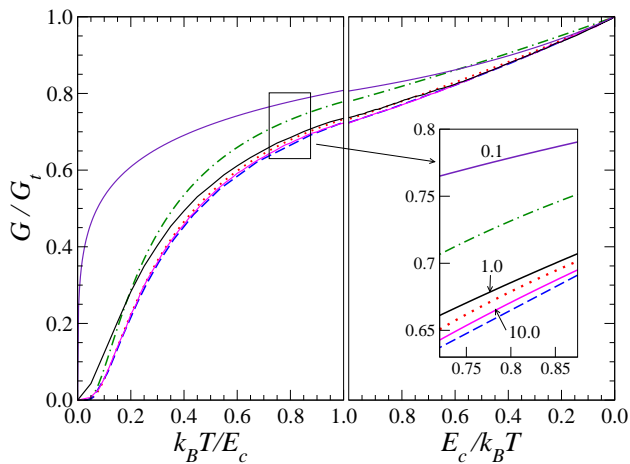}
\end{center}
\caption{
Conductance $G/G_t$ versus temperature.
The full lines are the numerical data from Ref. \cite{JoyezPRB},
obtained for $R/R_K = 0.1$, $1$, and $10$. The dashed line shows our
result \eqref{conductivity1} with $I(\alpha)$ calculated numerically
and the dotted line shows our
analytical approximation \eqref{conductivity2}.
The leading dependence $G/G_t = \exp{(-E_c/4k_BT)}$ is shown in a dotted-dashed line.
}
\label{FIGURE3}
\end{minipage}
\end{figure}

Moreover, we see that results \eqref{conductivity2} and \eqref{conductivity1} 
agree quite well also with the numerical data \cite{JoyezPRB} for $R/R_K = 1$. 
This is very surprising because $R = R_K$ contradicts the assumption $R \gg R_K$. 
Moreover, if we set $R = R_K$ into the assumption $(R_K/\pi R)E_c \ll k_BT$, 
we find that a good agreement can be expected only for $k_BT \gg E_c/\pi$.
However, one can see a good quantitative agreement down to temperature as 
low as $k_BT \simeq 0.15 E_c \simeq (R_K/2\pi R)E_c$ where the difference 
is about thirty percent (an acceptable error with regards to the decay by a factor of $\sim 6.5$). 
Moreover, one sees a qualitative accord for all temperatures.
 
We conclude that the restrictions $(R_K/\pi R)E_c \ll k_BT$ and $R \gg R_K$ 
can be soften to $(R_K/2\pi R)E_c \lesssim k_BT$
and $R \gtrsim R_K$, respectively. 
This extends the validity of the formulae \eqref{conductivity2} and \eqref{conductivity1} 
to a rather broad range of temperatures and series resistances, 
including the values $R \simeq R_K$ encountered in practice \cite{JoyezPRL}.

Why the formulae work so well? It seems that the conductance is rather robust against
the approximation \eqref{PEorthodoxtheoryGaussian}.
In this paper the robustness follows from comparison with the numerical data \cite{JoyezPRB} 
not restricted by approximation \eqref{PEorthodoxtheoryGaussian} 
and we do not attempt to give an analytical explanation. 
Note in figure 3 that even the leading dependence $G/G_t = \exp{(-E_c/4k_BT)}$ alone 
is able to capture the main trend both for $R/R_K = 1$ and $R/R_K = 10$.

Finally, we see in figure 3 that the formulae \eqref{conductivity1}, \eqref{conductivityLeadingdep},
and \eqref{conductivity2} fail to fit the numerical data \cite{JoyezPRB}
for $R/R_K = 0.1$. 
In this case $R$ is simply too low and even the soft restriction $k_BT \gtrsim (R_K/\pi R)E_c$ 
is fulfilled only for very high $T$.

In figure 4, formulae \eqref{conductivityLeadingdep} and \eqref{conductivity2}
are compared with experimental data \cite{JoyezPRL} for two samples and with
theory \cite{JoyezPRL} which holds for arbitrary $G_t$ and arbitrary $R$.
It was reported \cite{JoyezPRL} that $G^{-1}_t/R_K \simeq 1/3.07$
 for sample ($1$) and $G^{-1}_t/R_K \simeq 1/22.8$ for sample
($2$). Moreover \cite{JoyezPRL}, both samples posses the series resistance $R \simeq R_K$.
Clearly, formulae \eqref{conductivityLeadingdep} and \eqref{conductivity2}
fail to fit the data for sample ($2$) because the sample is too far 
from regime $G_t^{-1} \gg R \gg R_K$.
Sample ($1$) is more suitable in this respect and we expect a better fit. 
Indeed, both formulae mimic the S-shaped
character of the data for sample ($1$) and even fit the data as $T$ increases towards
$E_c/k_B$ and above.
This is a reasonable experimental support that $G \propto \exp{(-E_c/4k_BT)}$ 
is the leading dependence in a broad  temperature range. 
The discrepancy seen at low temperatures deserves a remark.

One source of the discrepancy is that also the sample ($1$) does not obey the condition $G_t^{-1} \gg R_K$ 
and hardly obeys even the softer version \cite{Heikilla} $G_t^{-1} \gg R_K/2 \pi$.
However, overall trend of the presented data suggests that the discrepancy would become much smaller 
if the experimental value of $G_t^{-1}$
is enhanced only a few times. Thus, one can soften $G_t^{-1} \gg R_K$ to $G_t^{-1} \gtrsim R_K$, 
or to $G_t^{-1} \gg R_K/2 \pi$ \cite{Heikilla}.
Another source of the discrepancy is that a sample with $R \simeq R_K$ does not fulfill well the 
condition $(R_K/2\pi R)E_c \lesssim k_BT$ as $T$ becomes low. 
It seems that the discrepancy would mostly disappear if one also enhances 
a few times the experimental value of $R$. 
Such sample would however still not fulfill the restriction $G_t^{-1} \gg R$ which suggests 
that this restriction is too stringent. This point needs a further investigation.

Inserting into the result $G/G_t=\exp{(-\alpha/4)}I(\alpha)$ the expansion
$\exp{(-\alpha/4)} \simeq  1- \alpha/4+\alpha^2/32 +O\left(\alpha^3\right)$
and expansion \eqref{correction2 small}, we obtain
the small $\alpha$ expansion
\begin{equation}
G/G_t \simeq 1-\alpha/3+\alpha^2/15+O\left(\alpha^3\right) \ ,
\label{conductivityexpansion for smal alfa}
\end{equation}
derived in Ref. \cite{JoyezPRB}
by a different method (equation (9) in Ref. \cite{JoyezPRB}, taken for
$R \gg R_K$).
Similarly, the path-integral study \cite{Wang} valid for junctions with arbitrary $G_t$ and $R$
reported a result which reproduces result \eqref{conductivityexpansion for smal alfa}
for $G^{-1}_t \gg R \gg R_K$.
Unlike Refs. \cite{JoyezPRB,Wang}, our derivation of result
\eqref{conductivityexpansion for smal alfa} relies on the knowledge of the
leading dependence $G/G_t \simeq \exp{(-\alpha/4)}$.

\begin{figure} [!t]
\centering
\includegraphics[width=8cm]{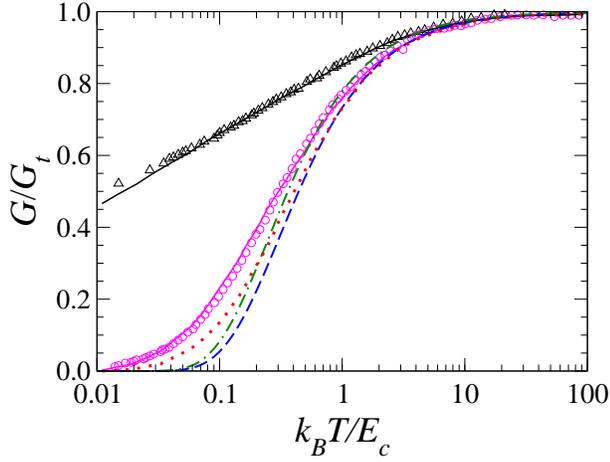}
\caption{
Conductance $G/G_t$ versus temperature.
Our formulae \eqref{conductivityLeadingdep} and
\eqref{conductivity2} are shown in a dotted-dashed line and in a dashed line.
The circles and triangles are the experimental data \cite{JoyezPRL} for samples
($1$) and ($2$) which posses different values of $G_t$
and $R$ (see the text).
The full lines are the numerical results \cite{JoyezPRL} of the theory valid for an
arbitrary $G_t$ and $R$. The numerical data
\cite{JoyezPRL} of the theory \cite{DevoretPRL,JoyezPRB} valid for small $G_t$ and arbitrary $R$
are shown in a dotted line.
These data are the same for both samples.}
\label{FIGURE4}
\end{figure}

\begin{figure} [!t]
\centering
\includegraphics[width=8cm]{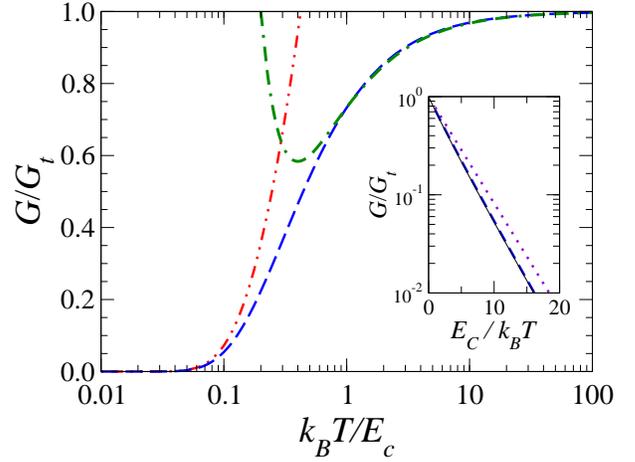}
\caption{
Conductance $G/G_t$ versus temperature. 
Formulae \eqref{conductivity2}, \eqref{conductivityexpansion for smal alfa},
and \eqref{largealphalimit} are plotted in a dashed line, in a dashed-double-dotted line,
and in a dotted-double-dashed line, respectively.
Inset: Arhenius plot $G/G_t$ versus $E_c/k_BT$.
The dashed line and dotted line show the formulae
\eqref{conductivity2} and \eqref{conductivityLeadingdep}, respectively.
The full line shows the numerical data \cite{GirvinPRL} obtained by solving
equations \eqref{Jtorthodoxtheory}, \eqref{PEorthodoxtheory}, \eqref{IVexpressionviaPE},
and \eqref{Kirchhoflaw1} for $R \rightarrow \infty$. 
These data were taken from the inset in figure 3 of Ref. \cite{GirvinPRL},
they essentially coincide with the numerical data 
\cite{JoyezPRB} for $R/R_K = 10$, shown in Fig. 3.}
\label{FIGURE5}
\end{figure}

If $\alpha \gg 1$, $I(\alpha) \simeq \sqrt{\pi^3/4\alpha}$ and equation
\eqref{conductivity1} gives
\begin{equation}
\frac{G}{G_t} \simeq  \sqrt{\frac{\pi^3}{4\alpha}}
\exp{\left(-\frac{\alpha}{4}\right)}  \ ,  \ \ \ \ \ \ \ \
\frac{R_K}{\pi R} \ll \frac{1}{\alpha} \ll 1 \ ,
\label{largealphalimit}
\end{equation}
which coincides with the known result \cite{Averin,AverinBook} for $\alpha^{-1} \ll 1$,
derived by a different approach.
Also the equation \eqref{largealphalimit} predicts the
leading dependence $G/G_t \simeq \exp{(-\alpha/4)}$,
however, only for small $\alpha^{-1}$ (for $k_BT \ll E_c$).
The fact that the leading dependence $G/G_t \simeq \exp{(-\alpha/4)}$
holds (with restriction $1/\alpha \gg R_K/\pi R$) for any $\alpha$,
can be concluded only from the exact result \eqref{conductivity1}. 
Moreover, as we have shown, restriction $R_K/\pi R \ll 1/\alpha$ 
should be soften to $R_K/2\pi R \lesssim 1/\alpha$.
Equation \eqref{largealphalimit} with restriction $R_K/\pi R \ll 1/\alpha \ll 1$ 
is also given in Ref. \cite{Odintsov}, but the leading dependence has a different pre-factor.

It may seem that results \eqref{conductivity1}, \eqref{conductivityLeadingdep},
and \eqref{conductivity2} do not differ very much from result \eqref{largealphalimit}.
The difference is however essential.

In figure 5 the formulae \eqref{largealphalimit} and \eqref{conductivityexpansion for smal alfa}
are plotted together with our result \eqref{conductivity2}. Unlike the result \eqref{conductivity2},
they cannot mimic the S-shaped experimental curve in the preceding figure
because they do not follow  the leading dependence $G \propto e^{-E_c/4k_BT}$
in an essential temperature range. Let us mention that it does not help very much if
we improve the formula \eqref{largealphalimit} so that we replace the factor $\sqrt{\pi^3/4\alpha}$
by a more precise expansion \eqref{correction big22}.

The fact that the leading dependence $G \propto e^{-E_c/4k_BT}$ exists
in a broad temperature range has been hidden in previous numerical 
studies \cite{GirvinPRL,JoyezPRL,JoyezPRB,Flensberg}.
Specifically, a strong, essentially exponential (Arhenius)
temperature  dependence was found numerically \cite{GirvinPRL,Flensberg} 
by solving equations \eqref{Jtorthodoxtheory},
\eqref{PEorthodoxtheory}, \eqref{IVexpressionviaPE},
and \eqref{Kirchhoflaw1} for $R \rightarrow \infty$.
In inset to figure 5, the Arhenius plot $G/G_t$ versus $E_c/k_BT$ found numerically in
Ref. \cite{GirvinPRL} is compared  with the formula \eqref{conductivity2} 
and dependence $G/G_t = e^{-E_c/4k_BT}$.
Obviously, the (almost) linear slope of the numerically 
generated Arhenius plot deviates from the slope $-E_c/4$.
To recognize that the numerical data follow the 
leading dependence $G \propto e^{-E_c/4k_BT}$,
one has to know result \eqref{conductivity2} or result \eqref{conductivity1}.

If one sets the probability \eqref{PEorthodoxtheoryGaussian}
into the equation \eqref{IVexpressionviaPE} and assumes $T \rightarrow 0$,
one finds analytically the well know \cite{DevoretPRL,GirvinPRL,GerdSchoen}
$I(V)$ characteristics
$I = \frac{G_t}{e} (eV-E_C) \Theta(eV-E_c)$,
where $\Theta(x)$ is the Heaviside step function.
This $I(V)$ result shows the Coulomb gap of size $E_C$ at zero temperature.
Noteworthy, the zero-voltage conductance $G \propto e^{-E_c/4k_BT}$ [results  
\eqref{conductivity1}, \eqref{conductivityLeadingdep},
and \eqref{conductivity2}] shows the effective gap $E_C/4$ at any temperature,
in reasonable accord with numerical studies and experiment.

Finally,
regime $k_BT \gtrsim \hbar /RC$ 
studied here should not be confused with the opposite
regime $k_BT \ll \hbar/RC$.
In the latter regime the quantum noise takes place instead 
of the semiclassical Nyquist noise 
and the junction conductance shows \cite{IngoldNazarov,GerdSchoen,Panyukov,SafiSauler} 
the power law behavior $G/G_t \sim [(\pi R/R_K)(2k_BT/E_c)]^{(2R/R_K)}$.

\section{IV. Summary and concluding remarks}

We have revisited the Coulomb blockade of the weak tunnel junction. 
We have analyzed the junction with capacitance $C$, connected in series 
with resistance $R$ which produces the Nyquist noise. 
In the semiclassical limit $k_BT \gg (R_K/2\pi R)E_c$ and $R \gg R_K$,
the Nyquist noise charges the junction by a random charge with a Gaussian distribution. 
Assuming the Gaussian distribution, we have derived analytically the temperature-dependent 
junction conductance $G(T)$ which works surprisingly well for temperatures
$k_BT \gtrsim (R_K/2\pi R)E_c$ and resistances $R \gtrsim R_K$.
Our result shows the leading
dependence $G(T) \propto e^{-E_c/4k_BT}$, so far believed \cite{Averin,Odintsov} 
to exist only in the limits $(R_K/\pi R)E_c \ll k_BT \ll E_c$ and $R \gg R_K$. 
The validity of our result for $k_BT \gtrsim (R_K/2\pi R)E_c$ and $R \gtrsim R_K$ 
has been confirmed by a good agreement with numerical studies \cite{GirvinPRL,JoyezPRL,JoyezPRB,Flensberg} 
which do no rely on the semiclassical limit, 
and by a reasonable agreement with experimental data  \cite{JoyezPRL} for $R$ as low as $R_K$.
Finally, our result reproduces the known \cite{Averin,AverinBook,Wang,JoyezPRB} 
asymptotic formulae valid in limits $T \ll e^2/2Ck_B$ and $T \gg e^2/2Ck_B$.

Our result can be useful in simulations of the nano-scale devices which involve 
a nano-scale tunnel junction.
Such junction is present say in a nano-scale field effect transistor \cite{roadmap}
or in a nano-scale memristor \cite{Zhirnov,Waser}.

\section{acknowledgement}
This work was supported by grant VEGA 2/0200/14 and by Structural Funds
of the European Union via the Research Agency of the Ministry of Education,
Science, Research and Sport of the Slovak republic,
project "CENTE II" ITMS code 26240120019.

\section{Appendix: Nyquist noise and distribution (\ref{PEorthodoxtheoryGaussian}) from semiclassical transport model}

Consider a metallic resistor with rectangular cross-section $L_y \times L_z$ 
and with length $L_x$ directed along the $x$ axis.
Assume that the resistor is in thermodynamic equilibrium.
The instantaneous electron current $i$, 
flowing in thermodynamic equilibrium in the $x$ direction, can be expressed as

\begin{equation}
i = \sum_{\textit{\textbf{k}},s} \frac{ev_x}{L_x} n(\textit{\textbf{k}},s) \ ,
\label{equlibriumcurrent}
\end{equation}
where $\textit{\textbf{k}}$ is the electron momentum, 
$s$ is the electron spin, $n(\textit{\textbf{k}},s)$
is the occupation number of state $(\textit{\textbf{k}},s)$, $v_x = \hbar k_x/m$ is the electron velocity,
and $m$ is the effective mass. 
The value of $n(\textit{\textbf{k}},s)$ fluctuates \cite{Kittel} around the mean value
$\langle n(\textit{\textbf{k}},s) \rangle = 1/(\exp{[(\epsilon(\textit{\textbf{k}}) - \mu)/k_BT]} + 1)$
with variance
\begin{equation}
\langle n^2 \rangle -  \langle n \rangle ^2 =  k_BT \frac{d}{d\mu} \langle n \rangle \ ,
\label{equlibriumfluctuations}
\end{equation}
where $\epsilon(\textit{\textbf{k}}) = \hbar^2 k^2/2m$ is the electron energy
at the orbital $(\textit{\textbf{k}},s)$, $\mu$  is the chemical potential,
and $\langle \dots \rangle$ is the ensemble average over the grand-canonical ensemble.
Therefore, the equilibrium current $i$ fluctuates around the mean value
\begin{equation}
\langle i \rangle = \sum_{\textit{\textbf{k}},s} \frac{ev_x}{L_x} \langle n(\textit{\textbf{k}},s) \rangle = 0
\label{meanequlibriumcurrent}
\end{equation}
with variance
\begin{equation} \begin{split}
\langle i^2 \rangle - \langle i \rangle^2 = \langle i^2 \rangle
= \frac{e^2}{L^2_x} \sum_{\textit{\textbf{k}},s} \sum_{\textit{\textbf{k}}^{\bf{,}},s^{,}} v_x v^{,}_x    &  \\
\
\times [\langle n(\textit{\textbf{k}},s) n(\textit{\textbf{k}}^{\bf{,}},s^{,}) \rangle -
\langle n(\textit{\textbf{k}},s) \rangle \langle n(\textit{\textbf{k}}^{\bf{,}},s^{,})\rangle]\ .
\end{split}
\label{varianceequlibriumcurrent}
\end{equation}
Assuming that each orbital is an independent grand-canonical system, we can use the equation
\begin{equation} \begin{split}
\langle n(\textit{\textbf{k}},s) n(\textit{\textbf{k}}^{\bf{,}},s^{,}) \rangle -
\langle n(\textit{\textbf{k}},s) \rangle \langle n(\textit{\textbf{k}}^{\bf{,}},s^{,}) \rangle  &  \\
\
= [\langle n^2(\textit{\textbf{k}},s) \rangle - \langle n(\textit{\textbf{k}},s) \rangle^2]
\delta_{\textit{\textbf{k}},\textit{\textbf{k}}^{\bf{,}}}\delta_{s,s^{,}}.
\end{split}
\label{uncorrelated}
\end{equation}
If the last equation is combined with equation \eqref{equlibriumfluctuations}
and then inserted into the equation \eqref{varianceequlibriumcurrent}, we find
\begin{equation}
\langle i^2 \rangle = k_BT \frac{e^2}{L^2_x} \sum_{\textit{\textbf{k}},s} v^2_x \frac{d}{d \mu} \langle n \rangle \ .
\label{variance2equlibriumcurrent}
\end{equation}
Using $\sum_{\textit{\textbf{k}},s} \rightarrow 2 \frac{L_xL_yL_z}{(2 \pi)^3} \int d\textit{\textbf{k}}$
and assuming $\mu \simeq E_F$ and
$d \langle n \rangle /d \mu \simeq d \langle n \rangle/d E_F \simeq \delta(\epsilon(k) - E_F)$ we get
\begin{equation}
\langle i^2 \rangle
= k_BT \frac{L_yL_z}{L_x} \frac{e^2 n_e}{m},
\label{variance2semifinal}
\end{equation}
where $n_e = {[(2m/\hbar^2)E_F]}^{3/2}/3 \pi^2$ is the electron density.
Expressing the resistor resistance $R$ via the Drude formula
$R^{-1}
= (e^2n_e\tau/m) (L_yL_z/L_x)$,
where $\tau$ is the electron momentum relaxation time, we can write  equation \eqref{variance2semifinal} as
\begin{equation}
\langle i^2 \rangle R
= k_BT \frac{1}{\tau} \ ,
\label{variance2final}
\end{equation}
where $\langle i^2 \rangle R$ is the mean electric power lost by the 
fluctuating current into the thermal reservoir.
The same power is delivered from the thermal reservoir to the electron gas.

Equation \eqref{variance2final} is an alternative form of the Nyquist result $\langle i^2 \rangle R
= k_BT \Delta f$, where $R$ is the resistance of the resistor connected to a lossless transmission line,
$\Delta f$ is the frequency bandwidth of the line, and $\langle i^2 \rangle R$ is the frequency
spectrum of the power irradiated by the electromagnetic waves propagating along the line (see page $102$
in Ref. \cite{Kittel}). The form \eqref{variance2final} follows from the transport-based considerations
which will be useful in what follows.

If the resistor is connected to the junction capacitor as in figure 1b,
the fluctuating current $i$ charges the capacitor by a certain charge $Q(t)$,
where $t$ is the charging time. Since each electron is scattered out of its orbital to another one
within the time $\tau$, the current $i$ changes its sign at random after each time step $\tau$ on average.
Due to the fluctuations of $i$, $Q$ fluctuates around the mean $\langle Q \rangle = 0$
and we search for the probability $P(Q) \ dQ$ that the capacitor is charged by a certain charge $Q$.
The charging of the capacitor with capacitance $C$ is governed by equation
\begin{equation}
\frac{d}{dt}Q(t) = i(t) - \frac{Q(t)}{RC}  \ .
\label{chargingeqaution}
\end{equation}
We multiply  equation \eqref{chargingeqaution} by $Q(t)^{2n-1}$, 
where $n =1$, $2$, $\dots$, and we perform the ensemble averaging  $\langle \dots \rangle$. 
We get
\begin{equation}
\frac{1}{2n} \frac{d}{dt} \langle Q(t)^{2n} \rangle = \langle Q(t)^{2n-1}i(t)\rangle - 
\frac{\langle Q(t)^{2n} \rangle }{RC} .
\label{averagedchargingeqaution}
\end{equation}
In steady state (for $t \gtrsim RC$) the left hand side of equation \eqref{averagedchargingeqaution} 
is zero and the right hand side does not depend on $t$. 
Keeping the $t$ dependence for formal reasons we have
\begin{equation}
\langle Q(t)^{2n} \rangle = RC \langle Q(t)^{2n-1}i(t)\rangle , \ \ \ n = 1, 2, \dots \ .
\label{averagedchargingeqaution2}
\end{equation}
We introduce the correlation function $\langle i(t')i(t) \rangle$ 
in a simple model in which $i$ is either $+ {\langle i^2 \rangle}^{1/2}$ or 
$- {\langle i^2 \rangle}^{1/2}$, chosen at random with the time step $\tau$. 
Then $\langle i(t')i(t) \rangle = 0$ for $ | t - t' | \geq \tau$ and 
$\langle i(t')i(t) \rangle = \langle i^2 \rangle $ for $ | t - t' | < \tau $, 
and additionally, $\langle Q(t')^{2n-1}i(t) \rangle = 0$ for $ | t - t' | \geq \tau $. 
An exponentially decaying correlation would not change our results.
We split the term $\langle Q(t)^{2n-1}i(t)\rangle$ in equation \eqref{averagedchargingeqaution2} 
into the terms
$ \langle Q(t-\tau)^{2n-1}i(t)\rangle $ and 
$ \int_{t-\tau}^{t} dt' \langle \frac{d}{dt'}  Q(t')^{2n-1} i(t) \rangle$, where
$\langle Q(t-\tau)^{2n-1}i(t) \rangle = 0$ as mentioned above. 
Thus
\begin{equation} \begin{split}
& \langle Q(t)^{2n-1}i(t)\rangle
\\
&
= (2n-1) \int_{t-\tau}^{t} dt' \langle Q(t')^{2n-2} \frac{d}{dt'}  Q(t') i(t) \rangle
\ .
\end{split}
\label{thetermontheright11}
\end{equation}
To calculate the integral on the right hand side of equation \eqref{thetermontheright11}, 
we insert for $dQ(t')/dt'$ the equation \eqref{chargingeqaution}. 
Then
\begin{equation}
\begin{split}
& \int_{t-\tau}^{t} dt'  \langle Q(t')^{2n-2} \frac{d}{dt'}  Q(t') i(t) \rangle  =
\\
&
\int_{t-\tau}^{t} dt'  \langle Q(t')^{2n-2} i(t') i(t)\rangle  - \int_{t-\tau}^{t} dt' \frac{ \langle  Q(t')^{2n-1} i(t)\rangle}{RC} \\
&
\simeq \ \ \tau \langle Q(t)^{2n-2} i(t)^2 \rangle - \frac{\tau}{RC} \langle  Q(t)^{2n-1} i(t)\rangle
\ ,
\end{split}
\label{thetermontheright3}
\end{equation}
where we have used relation $Q(t') \simeq Q(t)$ and after that relation 
$ \langle i(t') i(t)\rangle =  \langle i(t)^2 \rangle $. 
The latter relation is the correlation function introduced above and 
the relation $Q(t') \simeq Q(t)$ holds because for $ |t - t' | < \tau$
and $t > RC \gg \tau$ 
one typically has $|Q(t) - Q(t')|  \lesssim  \sqrt{\langle  i^2 \rangle } \tau \ll |Q(t)|$.

We insert equation \eqref{thetermontheright3} into the equation \eqref{thetermontheright11} and
we skip the term $- (\tau /RC) \langle  Q(t)^{2n-1} i(t)\rangle $ because $\tau \ll RC$. 
We obtain $\langle Q(t)^{2n-1} i(t)\rangle = (2n-1) \tau \langle  Q(t)^{2n-2}  i(t)^2 \rangle$. 
Combining the last equation
with equation \eqref{averagedchargingeqaution2} we find
\begin{equation}
\langle Q^{2n} \rangle = (2n-1) RC \tau \langle Q^{2n-2} i^2 \rangle , \ \ \ n = 1, 2, \dots \ ,
\label{averagedchargingeqaution44}
\end{equation}
where we have skipped the argument $t$. For $n=1$
\begin{equation}
\langle Q^{2} \rangle = RC \tau \langle i^2 \rangle .
\label{averagedchargingeqaution4}
\end{equation}
Further, $\langle Q^{2n-2} i^2 \rangle = \langle Q^{2n-2} \rangle \langle i^2 \rangle$ 
at least in the model in which
$i$ is either $+ {\langle i^2 \rangle}^{1/2}$ or $- {\langle i^2 \rangle}^{1/2}$.
So we finally have
\begin{equation}
\langle Q^{2n} \rangle = (2n-1) \langle Q^{2n-2} \rangle \langle Q^2 \rangle , \ \ \ n = 1, 2, \dots \ .
\label{averagedchargingeqaution5}
\end{equation}
Obviously, the only distribution which gives the moments $\langle Q^{2n} \rangle = \int dQ  Q^{2n} P(Q)$ 
that fulfill the equation
\eqref{averagedchargingeqaution5} for all $n$ is the Gaussian distribution
\begin{equation}
P(Q) \ dQ = \frac{1}{\sqrt{2\pi \langle Q^2 \rangle}} 
\exp{\left[-\frac{Q^2}{2\langle Q^2 \rangle}\right]} \ dQ \ .
\label{distributionPQ}
\end{equation}
Taking into account equations \eqref{averagedchargingeqaution4} and \eqref{variance2final}, 
we obtain
\begin{equation}
P(Q) \ dQ = \frac{1}{\sqrt{2\pi k_BT C}} \exp{\left[-\frac{Q^2}{2k_BTC}\right]} \ dQ \  .
\label{distributionPQtimeRC}
\end{equation}
Finally, each tunneling event is accompanied by environmental excitation with 
energy \cite{GerdSchoen,IngoldHabilitation,Heikilla}
\begin{equation}
E = \frac{(Q \pm e)^2}{2C} - \frac{Q^2}{2C} = \pm \frac{eQ}{C} + \frac{e^2}{2C} \,
\label{excitationenergy}
\end{equation}
where $Q^2/2C$ and $(Q \pm e)^2/2C$ are the electrostatic energies of the junction 
before and after the tunneling, respectively, 
and the sign (plus or minus) depends on whether the tunneling charges or discharges the junction.
We can use equation \eqref{excitationenergy} and replace the variable $Q$ in the distribution 
\eqref{distributionPQtimeRC} by variable $E$. 
We readily obtain the distribution \eqref{PEorthodoxtheoryGaussian}.

From the above derivation one sees that the distribution \eqref{distributionPQtimeRC} 
is a steady state result,  valid for time $t \gtrsim RC$. 
This distribution can be meaningfully used in the $I(V)$ calculation of Sect. II 
only if $RC \ll e/I$, where $e/I$ is the time between two subsequent tunneling events. 
Condition $RC \ll e/I$ also justifies  the concept of the ideal 
voltage source \cite{Cleland} assumed in figure 1b.

Finally, the above derivation is semiclassical in the sense that it uses 
the quantum occupation numbers but the equation of motion for $Q(t)$ is classical. 
A profound quantum treatment is given in Ref. \cite{Frey}. 
According to the quantum theory reviewed in Sect. II the semiclassical approach 
is valid if $\hbar /RC \ll k_BT$. 
However, we find in Sect. III that the restriction
$\hbar /RC \ll k_BT$ should be soften to $\hbar /2RC \lesssim k_BT$ in the conductance case.

\end{document}